\magnification=1200
\input epsf

\vsize=22truecm
\hsize=14truecm
\baselineskip=20pt
\parskip=0.2truecm
\hoffset=1.5truecm
\headline{\hfil\tenrm\folio\hfil}
\footline={\hfil}
\headline{\ifnum\pageno=1\nopagenumbers
\else\hss\number\pageno\fi}
\footline={\hfil}
\baselineskip=14pt
\vbox to 1,5truecm{
}
\centerline{\bf ORSAY LECTURES ON CONFINEMENT (II)}\bigskip
\centerline{by}\smallskip
\centerline{ {\bf Vladimir N. Gribov}\footnote{*}{Supported in part by Landau
Institute-ENS D\'epartement de Physique exchange program}}
\centerline{L. D. Landau Institute for Theoretical Physics} \par
\centerline{Acad. of Sciences of the USSR, Leninsky pr. 53, 117 924 Moscow,
Russia}
\par \medskip \centerline{and} \medskip
\centerline{KFKI Research Institute for Particle and Nuclear Physics} \par
\centerline{of the Hungarian Academy of Sciences}\par
\centerline{H-1525 Budapest 114, P.O.B. 49, Hungary}\medskip
\centerline{and} \medskip
\centerline{Laboratoire de Physique Th\'eorique et Hautes Energies%
\footnote{**}{Laboratoire associ\'e au Centre National de la Recherche
Scientifique} }
\centerline{Universit\'e de Paris XI, b\^atiment 211, 91405 Orsay Cedex,
France}\medskip
\vskip 6 cm
\noindent LPTHE Orsay 94-20 \par
\noindent February 1994
\vfill \supereject

\baselineskip=20pt
\noindent {\bf  \underbar {The confinement of the heavy quark}}\medskip

In the previous talk [1] we have considered some aspects of the theory of
supercharged
nuclei, and came to the conclusion, that the superbound atoms are stable mainly
due to
the Pauli principle.\par

Before going to the heavy quark, let us discuss briefly, how the problem of the
supercharged nucleus can be handled practically. One possibility to formalize
it is the
following. We have to calculate the Green function of the Dirac equation in
external
field~:

$$\widehat{\nabla}\ G(x, x') = {1 \over i} \delta(x - x') \ \ \ , \eqno(1)$$

\noindent where

$$\widehat{\nabla} = \gamma_{\mu} \left ( \partial_{\mu} - i A_{\mu} \right ) \
\ \ .
\eqno(2)$$

\noindent The initial external field is that of the considered nucleus.
However, in the
presence of a charge $Z$ the stationary states correspond to atomic states with
a charge
$Z-N$ where $N$ is large enough to fulfil $Z-N < Z_{cr}$, i.e. there have to be
$N$
electrons rotating around the nucleus. But if so, we will have to take into
account, that
the field which acts on each electron is also changing. \par

The ground state in our case is an atomic type state. This means, that although
we don't
know what our field is, we can find it. The gradient squared of the zeroth
component $A_0
\equiv A$ of the potential has to be equal

$$\nabla^2 A_0 = e_0^2 \left ( \rho_{nucl} + \rho_{el} \right ) \ \ \ .
\eqno(3)$$

\noindent where $\rho_{e\ell} = < \bar{\psi} \gamma_0 \psi >$. \par

The charge density will depend on the potential, and the equation becomes a
Thomas-Fermi
type equation for the potential existing in this system. If we find a wave
function
satisfying the equation we will know $A$, the Green function and $\rho$. This
means, that
the formal way of solution is just to solve a self-consistent, Thomas-Fermi
type equation
for the effective atomic potential. \par

Let us consider the total charge $Q$

$$Q = \int \rho d^2r \ \ \ ; \eqno(4)$$

\noindent the density is the average value

$$\rho = <\bar{\psi}(r) \psi(r)> = \sum_{\omega < 0} \bar{\psi}_n(r) \psi_n(r)
\eqno(5)$$

\noindent which in the Dirac picture is the sum of all negative energy levels.
The total
charge $Q$ will be the sum of the energy levels over all $\omega_n$. The sum is
divergent,
and we have to make a cut-off and subtract the bare particles. This, however,
will not be
enough. Indeed, if we just subtract the value of the charge of the free vacuum,
then at $Z
\not= 0$ the charge of the vacuum (i.e. of the nucleus) will not be equal $Z$
and will
continue to diverge logarithmically. We have to subtract the value at small $Z$
in a way
which gives zero for the total charge of the electron vacuum (this corresponds
to the
correct renormalization of the charge of the nucleus). \par

The above procedure does not reflect literally the subtraction of the vacuum
charge without
external field. This becomes especially obvious, if one makes use of the
Levinson theorem
which connects the number of states of a particle in external field in a given
energy
interval with the phase of scattering of this particle in the same field. The
number $N$ of
additional states in the energy interval between $E_1$ and $E_2$ is defined by
the
difference of phase shift

$$N = {1 \over \pi} \left [ \delta(E_2) - \delta(E_1) \right ] \ \ \ .
\eqno(6)$$

\noindent Because of this, the number of new states in the Dirac vacuum equals

$$N = {1 \over \pi} \left [ \delta(- m) - \delta(- \infty) \right ] \ \ \ .
\eqno(7)$$

\noindent For $Z < Z_c$ this number should be zero. However, for the Dirac
equation in
external field this condition is not fulfilled, because $\delta(- \infty) \not
= 0$, and in
order to obtain the proper definition of the vacuum charge we have to subtract
a quantity
which, generally speaking, depends on $Z$ ($\delta(- \infty)$). This
subtraction means, in
fact, that we have to change the interaction with the external field when $E
\to - \infty$
so that $\delta(- \infty) = 0$. ($\delta(- m)$ can always be considered to be
zero if the
field is small). If, however, $Z$ will be increased and becomes more than
critical, then,
as we saw, the levels will pass the point $E = - m$ and move to the complex
plane. It is
easy to check, that every time this happens the phase $\delta(- m)$ is changing
by $\pi$ and
the move of $n$ levels into the complex plane changes the number of states by
$n$ so that
the charge of the vacuum (i.e. of the atom) becomes $Z - n$. The value of
$\rho_e$ which
enters the equation for the self-consistent field is to be defined by the
contribution of
levels which passed through $E = - m$. \par

We discussed in detail this concrete structure in order to refer to it in the
following,
talking about the heavy quark in the vacuum of the light quark. We shall
suppose, that
due to gluonic vacuum polarization the effective coupling $\alpha(r)$ which at
small $r$
has the usual perturbative behaviour, reaches a constant value at $r >> r_0$
($r_0 \sim
1/\lambda)$, and this constant value will be more than unity. (Without this
ansatz,
allowing that $\alpha$ continues to grow, things are more complicated, but
nothing
essentially changes).

$$
\epsfbox{II_1.ps}
$$\nobreak
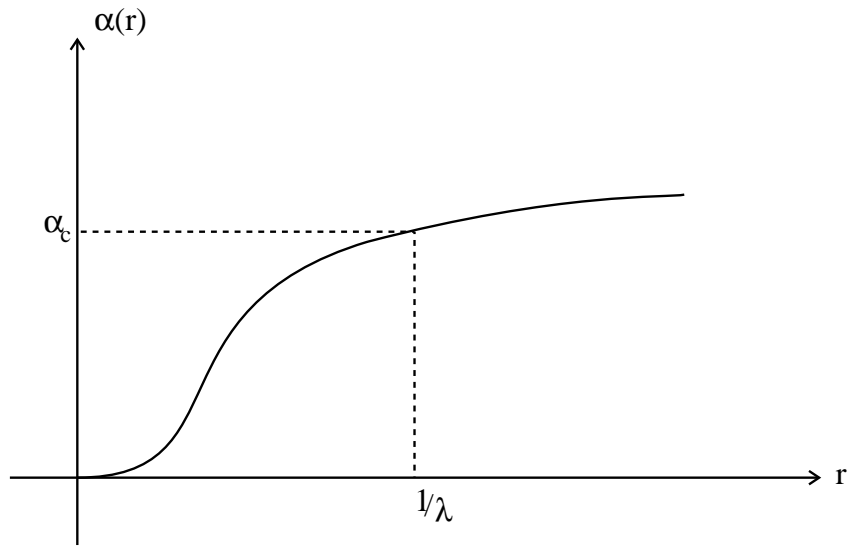
\centerline{Figure 1} \par
\centerline{The supposed behaviour of the effective coupling.} \par \smallskip

\noindent It is natural to expect such a behaviour of the charge in QCD. But
even an
Abelian theory can reveal such a behaviour of charge, if it originates from a
non-Abelian
theory via spontaneous colour symmetry breaking. In this case the charge will
increase,
and as a result, 6-7 gluons acquire masses. After that there remain one or two
massless
objects - ``photons'' - and the behaviour of charge will be exactly as
discussed, because
after the symmetry breaking we have $r_0 = {1 \over m}$ (heavy gluon). We can
ask now~:
what happens in the vacuum of light quarks under these circumstances~? Outside
the region
where $\alpha$ is growing, we will have an Abelian theory and we can consider
the quark
states in the normal way.

$$
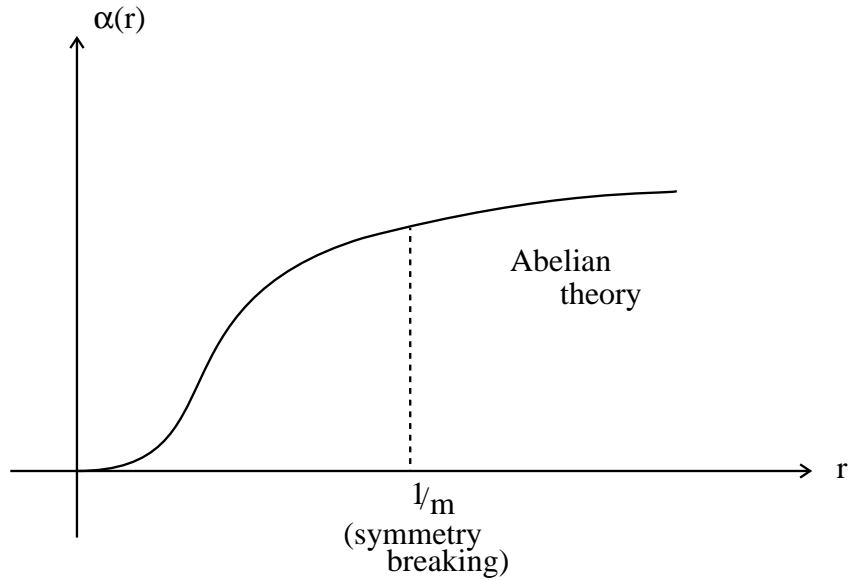
\epsfbox{II_2.ps}
$$

\noindent If $\alpha$ becomes more than critical, the corresponding level goes
to the Dirac
sea, which, consequently, will have to be filled up and we will find an atomic
type state.
If this is so, the charge density will be exactly the same as in the case of a
supercharged
nucleus. The $\lambda_{QCD}$ is analogous to the radius of the nucleus. There
is, however,
an important difference between QED and this case. Indeed, in QED we consider a
nucleus
with a charge $Z$ in the centre, and we put one or two electrons around it to
organize an
atom. In QCD we have a heavy quark with a very small intrinsic charge.

$$
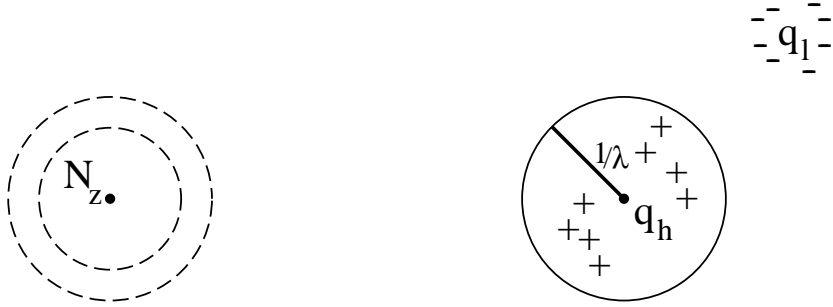
\epsfbox{II_3.ps}
$$

\noindent Due to vacuum polarization the charge becomes large inside
$1/\lambda$. The system
creates an empty level, and we have to fill it. This means, that we have to add
an intrinsic
charge, equal to that of the heavy quark, but with the opposite sign (outside
the region of
$1/\lambda$). So we have here two intrinsic charges with the total charge zero,
and of
course the vacuum polarization can never change the total charge. In other
words, in QCD
the supercritical atom would be a meson-type state with zero colour charge. Our
task is to
show, that these are not only words  - we have to include the mechanism of
vacuum
polarization formally. In order to do so, let us discuss the problem in a
language similar
to QED, neglecting non-Abelian fluctuations of the colour field $A$ - the
average field of
heavy and light quarks inside the meson. In this case $A_0$ is defined by the
equation

$$\nabla^2A = e_0^2 \left ( \rho_h - \rho_{\ell} \right ) \ \ \ , \eqno(8)$$

\noindent where $\rho_h$ is the density of the heavy quark, and $\rho_{\ell}$
that of the
light quark. Again,  we can write the equation for the Green function

$$\left ( \widehat{\nabla} - m \right ) G(x, x') = {1 \over i} \delta(x) \ \ \
. \eqno(9)$$

\noindent But now the problem is, how to calculate this. What means, e. g.,
$\rho_{heavy}$~?
We can not use the expression $\rho_h = e_0^2 \delta(r)$, since it does not
include the
bosonic vacuum polarization, which has to be taken into account. It is well
known, that in
principle vacuum polarization means summing a diagram like

$$
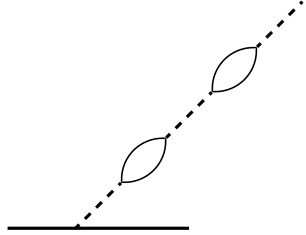
\epsfbox{II_4.ps}
$$

\noindent with gluonic loops inside. The problem is, however, how to write this
in normal
space-time language. There is a good way to sum this diagram by writing an
equation for
currents. Knowing the external current and wishing to calculate the total
current, we have
to solve the equation

$$j_{\mu}(x) = j_{\mu}^{ext}(x) + {\alpha_0 b \over 2 \pi^2} \int \delta '
\left ( (x - x')^2
\right ) j_{\mu}(x') d^4x' \ \ \ , \eqno(10)$$

\noindent which corresponds to the summation of the diagram. We will not derive
this
equation, because it is almost obvious for the static charge which we are
interested in.
\par

Let us consider a static charge, not depending on time. In this case

$$\rho(r) = \rho_{\mu}^{ext}(r) + {\alpha_0 b \over 8 \pi^2} \int {d^3r' \over
|r -
r'|^3} \rho(r') \ \ \ , \quad |r - r'| > \varepsilon \ \ \ , \eqno(11)$$

\noindent where $1/\varepsilon$ is the ultraviolet cut-off, $b = {11 \over 3}
n_c - {2
\over 3} n_f$. \par

The solution of this equation is very simple and leads to the usual expression
for charge
renormalization in QCD. In order to see this, let us introduce the quantity
$Q(r)$~:

$$Q(r) = \int_0^r \rho(r') 4 \pi r'^2 dr' \ \ \ . \eqno(12)$$

\noindent The logarithmic derivative of $Q(r)$ is

$$\partial_{\xi} Q(r) = r {\partial Q(r) \over \partial r} = 4 \pi r^3 \rho(r)
\ \ \ .
\eqno(13)$$

\noindent  For $Q(r)$ we can write the equation

$$\partial_{\xi} Q(r) \ =\ \partial_{\xi} Q_{ext}(r) + {\alpha_0 b \over 8
\pi^2} \int_{|r -
r'|> \varepsilon} {d^3r' \over |r - r'|^3} \ {r^3 \over r'^3} \partial_{\xi '}
Q(r') \ \ \ .
\eqno(14)$$

\noindent The integration in the right hand-side of this expression contains
two logarithmic
regions : $|r - r'| << r$ and $r >> r'$. In the first region, $\partial_{\xi '}
Q(r')$ can
be substituted by $\partial_{\xi} Q(r)$ ; integrating over the second region,
$r'$ in the
denominator $|r - r'|^3$ can be neglected. As a result, we obtain

$$\partial_{\xi} Q(r) = \partial_{\xi} \ Q_{ext}(r) + {\alpha_0 b \over 8
\pi^2}
\int_{\varepsilon}^r {d^3 r'' \over r''^3} \partial_{\xi} \ Q(r) + {\alpha_0 b
\over 2 \pi}
Q(r) \ \ \ ,$$

\noindent which is equivalent to

$$\partial_{\xi} \left ( 1 - {\alpha_0 b \over 2 \pi} \ell n {r \over
\varepsilon} \right )
Q(r) = \partial_{\xi} \ Q_{ext}(r) \eqno(15)$$

\noindent or

$$Q(r) = {Q_{ext}(r) \over 1 - {\alpha_0 b \over 2 \pi} \ell n {r \over
\varepsilon}} \ \ \ .
\eqno(16)$$

\noindent For a point-like charge

$$Q_{ext}(r) = 1 \ \ \ ,$$

\noindent and we have

$$A(r) = {\alpha (r) \over r} = \alpha (r) A_{ext} \ \ \ . \eqno(17)$$

\noindent The concrete expression

$$\alpha(r) = {\alpha_0 \over 1 - {\alpha_0 b \over 2 \pi} \ell n {r \over
\varepsilon}}
\eqno(18)$$

\noindent obtained from perturbation theory has, of course, an unphysical
singularity. For a
point-like charge (18) has to be substituted by an expression
cor\-res\-pon\-ding to the
behaviour of $\alpha(r)$ as shown in Fig. 1. However, for the distributed
charge the relation
between the external field and the field which takes into account the
polarization is
non-local in coordinate space. The correct expression for the relation between
the external
field and the observable field is local in the momentum space~:

$$A(q) = \alpha(q^2) A_{ext}(q) \eqno(19)$$

\noindent which leads in the coordinate space to an expression of the following
type~:

$$A(r) = \int K(r - r') A_{ext}(r') d^3 r' \ \ \ ,  \eqno(20)$$

\noindent where

$$K(r) = \int e^{iqr} \alpha(q) {d^3q \over (2 \pi)^3} \ \ \ .$$

\noindent Similarly,

$$\rho(r) = \int K(r - r') \rho_{ext}(r') d^3r' \ \ \ . \eqno(21)$$

\noindent If we now suppose, that $\alpha(q)$ as a function of $1/q$ behaves
according to
Fig. 1, and at large $q$ values it is defined by perturbation theory, then the
equation (8)
has to be understood as an equation for $A_{ext}$ :

$$\nabla^2 A_{ext}(r) = \delta(r) - \bar{\psi}_{\ell}(r) \gamma_0
\psi_{\ell}(r) \eqno(22)$$

\noindent where $\psi_{\ell}(r)$ is the solution of the Dirac equation

$$(H + A) \psi = E \psi$$

\noindent for the light antiquark in a superbound state in the field $A(r)$
defined by
equation (20). The solution of this problem gives the energy and the features
of the meson
$q_n \bar{q}_{\ell}$ with zero total charge. Due to (21),

$$Q = \int K(r) d^3r \ Q_{ext} = \alpha(q = 0) Q_{ext} \eqno(23)$$

\noindent with $Q = 0$ if $Q_{ext} = 0$. \par

We have just proved, that because of the big charge which appears through
vacuum
polarization, in the case of QCD the ``atomic'' bound state will, indeed, be a
meson. The
heavy quark will decay into a $q_h \bar{q}_{\ell}$-meson and a light quark :

$$q_h \to M \left ( q_h \bar{q}_{\ell} \right ) + q_{\ell} \ \ \ . \eqno(24)$$

In the next lecture we shall consider light quark states. So far there is one
important
thing to stress, namely if we don't include essential interactions between
light quarks in
the vacuum, we come to a reasonable conclusion for the case of the heavy quark
but, as we
will just see, to an unreasonable one about the light quark. Indeed, let us try
to extend
the considered procedure to the latter case. Suppose, that there is a light
quark moving,
and a potential acts on its vacuum. What will we see classically~? Since the
Coulomb field
is a vector field, it is shrinking, but the total integral remains the same.
Because of
this, we will find immediately, that there is a bound state in this potential
even for
fastly moving particles. This, however, means, that we have here an unstable
state, which
has to be filled, and as a consequence the light quark will decay into a meson
and a light
quark again~:

$$q_{\ell} \to M + q_{\ell}$$

\noindent which, of course, contradicts the energy conservation, unless the
appearing meson
is of negative energy. In order to have a self-consistent picture, we have to
suppose that
the light quark in the vacuum will interact so strongly that there have to be
negative
energy levels and the whole vacuum has to be rearranged. So from the picture we
described we
come quite naturally to light quark interactions. We will see, that these
interactions are,
indeed, very strong and lead to the confinement of light quarks which will take
place at
relatively small ${\alpha \over \pi}$ values ; this means, that the overall
corrections for
vacuum polarization will not be large. \par

This is for the future. What we have to add now, is, that even in the language
which was
accepted so far, with no strong interactions between particles in the vacuum,
the problem in
real QCD which is non-Abelian is more complicated. In QED we have one charge,
and all the
electrons in the vacuum interact with this charge, independently from each
other. In QCD
this can take place only, if the field of the heavy quark is an Abelian one. It
is highly
probable, that this is, indeed, the case, when the field of the heavy quark
becomes large as
a result of gluonic vacuum polarization.

\medskip
\noindent {\bf  \underbar {Reference:}}\medskip
[1] V.N. Gribov, Orsay Lectures on confinement (I)\hfill\break
\phantom{[1]} The theory of supercharged nucleus, LPTHE ORSAY 92/60
(1993).

\end